\documentstyle[11pt,epsfig,color]{article}
\begin{document}

\begin{center}
{\Large\bf A Reconstruction of Quintessence Dark Energy}
\\[15mm]
Ankan Mukherjee, \footnote{E-mail: ankan\_ju@iiserkol.ac.in}~~
Narayan Banerjee \footnote{E-mail: narayan@iiserkol.ac.in}

{\em Department of Physical Sciences,~~\\Indian Institute of Science Education and Research Kolkata,~~\\Mohanpur, West Bengal-741246, India}\\[15mm]
\end{center}

\vspace{0.5cm}
\vspace{0.5cm}
\pagestyle{myheadings}
\newcommand{\be}{\begin{equation}}
\newcommand{\ee}{\end{equation}}
\newcommand{\bea}{\begin{eqnarray}}
\newcommand{\eea}{\end{eqnarray}}

\begin{abstract}
 With a parametric form of the equation of state parameter of dark energy, a quintessence potential  has been reconstructed. The potential is found to be a generalization of a double exponential potential. The constraints on the parameters are obtained by maximum likelihood analysis using observational Hubble data, type Ia supernova data, Baryon Acoustic Oscillation data and the CMB shift parameter data. The model shows preference towards the phantom behaviour of dark energy.
\end{abstract}

\vskip 1.0cm



\section{Introduction}
Consistent with the indications given by the observations in the late nineties \cite{riess,perl}, subsequent observations \cite{barrisbj,knopra,tonryjl}  have confirmed that the universe at present is undergoing an accelerated phase of expansion. The observations also indicate that the alleged acceleration is rather a recent phenomenon \cite{tr} which came into being well within the matter dominated regime. The  component of the matter sector, responsible for  this acceleration, still eludes any observational detection or even a unique theoretical prediction. A cosmological constant certainly does very well in explaining this dynamics of the universe, but it has the huge discrepancy between the observed value and the theoretically predicted one. A scalar field, called the quintessence field, can certainly drive the acceleration with the aid of a suitably chosen potential, but no scalar potential has a firm theoretical support. There are excellent reviews that describe the suitability and problems of various models for an accelerated universe \cite{rev,copelandsami,martinj,padmanavant}. A modification of General Relativity (GR) is also looked at, leaving the matter sector intact. However, as GR describes the local astronomy so efficiently, the search for the matter component  responsible for the accelerated expansion, popularly dubbed as dark energy, is still very much alive  so as to keep the basic therory of gravitation unaltered.
\par One of the attempts towards finding a quintessence field is  ``reconstruction", i.e., building up the model from the observational data. This kind of ``reverse way" of finding a scalar potential has been there for a long time in the literature\cite{em}. The idea is to assume a particular evolution scenario, consistent with the observational requirement, and then to look for a matter field giving rise to that kind off evolution. In the context of dark energy, this method was initially utilised by Starobinsky \cite{saa} who used density perturbation and also by Huterer and  Turner \cite{huterer1,huterer2} where the data of distance measurement were invoked. In this context we also refer to the work by Saini {\it et al.} \cite{sainitd}.
\par In  the absence of a clear indication in favour of any particular potential, this kind of investigation has gained a lot of attention. Reconstruction of a dark energy potential normally involves the search for the equation of state parameter for scalar field, $ w_{DE}=\frac{p_{\phi}}{\rho_{\phi}}$ , where $p_{\phi}$ and $ \rho_{\phi}$ are the contributions of the scalar field to the pressure and density sectors respectively. A review on the initial attempts in this direction can be found in \cite{vs}. A reconstruction of $ w_{DE}$ may be accomplished in two ways. One is to choose a form of $ w_{DE}=w_{DE}(z)$, where $z$ is the redshift and estimate the parameters in $ w_{DE}$ with the help of the observational data \cite{gerke,gongwang,holsclawt} . The other is to directly find the functional form of $w_{DE}(z)$ from the data. Recently Holsclaw {\it et al} \cite{hu} discussed the second kind of reconstruction, a non-parametric one, as an inverse statistical problem. Sahl$\acute{e}$n {\it et al} also discussed a direct or non-parametric reconstruction of the quintessence potential \cite{slp,slp2}. The time evolution of $w_{DE}$ had been reconstructed very recently with non-parametric Bayesian method by Crittenden {\it et al} \cite{czp}. Pan and Alam investigated the usefulness of various cosmological parameters in selecting or rejecting different reconstructed dark energy models \cite{pa}. Nair {\it et al} adopted the Gaussian processes to explore the scalar field dynamics \cite{nair}. In the context of other theories of gravity, such as  the scalar tensor theories of gravity, reconstruction of dark energy has also been looked at by many \cite{boisseau,perivolaropolos,neupane,grandaln}. 
\par The present work deals with the former, that is, the parametric approach for a reconstruction of the quintessence potential. The reconstruction is based upon the parametrization  of dark energy  equation of state parameter $w_{DE}(z)$.  Different dark energy equation of state $w_{DE}(z)$ models have been constrained using the recent observational data sets by Xia, Li and Zhang \cite{XiaJQ}.  In this paper, a new parametric dark energy equation of state parameter has been proposed.  The statistical analysis of this model is carried out using the type Ia supernova  distance modulus data (SNeIa), observational Hubble data (OHD), Baryon Acoustic oscillation data (BAO) and the CMB shift parameter data (CMBShift). It is imperative to note that most of the $w_{DE}$ parametrizations slightly favour the phantom behaviour of dark energy, that is $w_{DE}<-1$. This model also shows preference towards the phantom behaviour  across the limit $w_{DE}=-1$. The paper has been arranged as the following. The mathematical framework, including the complete solution of the set of Friedmann equations, is given in section 2. Section 3 contains the results of the statistical analysis. Finally in section 4, we have concluded with an overall discussion regarding this dark energy model. 
    
\section{Reconstruction of the scalar field potential from the equation of state of the scalar field}
The field equations for a spatially flat FRW universe with cold dark matter (given by a pressureless fluid) and a scalar field are 
\be
3H^2=8\pi G(\rho_m + \rho_{\phi}),
\label{friedmann1}
\ee
\be
2\dot{H}+3H^2=-8\pi G p_{\phi},
\label{friedmann2}
\ee
where $H$ is the Hubble parameter given by $H=\frac{\dot{a}}{a}$ ($a$ being the scale factor), $\rho_m $ is the matter energy density and $ \rho_{\phi}$  and $p_{\phi}$ are the contributions of the scalar field to the energy density and pressure sectors respectively. The latter two are given by 

\be
\rho_{\phi}=\frac{\dot{\phi}^2}{2}+V(\phi),
\label{rhophi}
\ee
\be 
p_{\phi}=\frac{\dot{\phi}^2}{2}-V(\phi),
\label{pphi}
\ee
where $V(\phi)$ is the scalar potential. An overhead dot indicates a differentiation with respect to the cosmic time $t$. The pressureless cold dark matter satisfies its own conservation equation which leads to 
\be
\rho_m= \rho_{m0}(1+z)^3,
\label{rhom}
\ee
where $z$ is the redshift parameter defined as $1+z=\frac{a_0}{a}$ where $a$ is the scale factor and $a_0$ is its present value. This $\rho_{m0}$ is the present value of the dark matter density. With equations (\ref{friedmann1}),(\ref{friedmann2}) and (\ref{rhom}), the wave equation for the scalar field,
\be
\Box\phi+\frac{dV}{d\phi}=0,
\label{waveequn}
\ee 
is a consequence of the Bianchi identity and does not lead to an independent equation. From equations (\ref{friedmann1}) and (\ref{friedmann2}), the equation of state parameter $w_{DE}$ can be written as
\be
w_{DE}=\frac{p_{\phi}}{{\rho}_{\phi}}=-\frac{2\dot{H}+3H^2}{3H^2-8\pi G\rho_m}.
\label{wDE}
\ee
One can replace the argument '$t$' by the redshift $z$ in this equation. With the aid of the equation (\ref{rhom}), the equation (\ref{wDE}) would look like 
\be
2(1+z)H\frac{dH}{dz}=3(1+w_{DE})H^2-8\pi G\rho_{m0}(1+z)^3w_{DE}.
\label{differentialequn}
\ee  
As we have three unknown quantities  $a$, $\phi$ and $V(\phi)$ against only two equations, namely equation (\ref{friedmann1}) and (\ref{friedmann2}) to solve for them, we can choose an ansatz so as to close the system of equations. In what follows, a one parameter equation of state parameter, given by 
\be 
w_{DE}(z)=-\frac{3}{\alpha (1+z)^3+ 3}, 
\label{wz}
\ee  
 is chosen where $\alpha$ is  a constant parameter. The reason for choosing this kind of $w_{DE}$  is that for high $z$, i.e. at the early stage of evolution, $w_{DE}$ is almost zero so that it is hardly distinguishable from the equation of state parameter of a pressureless fluid, but  gradually decreases to more and more negative values so as to yield an increasing negative pressure. For $\alpha = 0$, $w_{DE}$ reduces to $-1$, i.e., that of a cosmological constant. For $\alpha<0$, the model leads to a phantom behaviour i.e. $w_{DE}<-1$. Normally a dark energy model is chosen such that it remains subdued at an early stage, i.e., for high value of $z$ and evolves to dominate over the dark matter only at the later stage of evolution. The present choice is qualitatively different from such models as at the early stage with $w_{DE}$ approaching zero, the dark energy is rather indistinguishable from the pressureless dark matter.

\par With equation (\ref{wz}), one can integrate equation (\ref{differentialequn}) to obtain 
\be 
H^2(z)=H_0^2\left[\frac{(\alpha+3\Omega_{m0})}{(\alpha+3)}(1+z)^3+\frac{3(1-\Omega_{m0})}{(\alpha +3)}\right], 
\label{Hz}
\ee
where $H_0$ is the present value of Hubble parameter and $\Omega_{m0}$ is the present density parameter given by $\Omega_{m0}=\frac{8\pi G\rho_{m0}}{3H_0^2}$. The deceleration parameter $q$, defined as ($-\frac{a\ddot{a}}{{\dot{a}^{2}}}$), can be written in terms of $z$ as
\be
q(z)=-1+\frac{3(\alpha+3\Omega_{m0})}{2(\alpha+3)}\frac{H_0^2(1+z)^3}{H^2(z)}.
\label{qz}
\ee

The nature of evolution  of $q(z)$ can be investigated utilizing the values of the parameters $\alpha$ and $\Omega_{m0}$, constrained by observation. Now from equation (\ref{friedmann1}) and (\ref{friedmann2}), one can write (using the expression for $ \rho_{\phi}$  and $p_{\phi}$) 
\be
2\dot{H}=-8\pi G(\rho_{m}+\dot{\phi}^2),
\label{hdot}
\ee
which can be written as
\be
8\pi G(1+z)^2H^2\Bigg(\frac{d\phi}{dz}\Bigg)^2=2(1+z)H\frac{dH}{dz}-3H_0^2\Omega_{m0}(1+z)^3,
\label{phiequation}
\ee
if $z$ is used as the argument instead of $t$. This can be integrated to yield (using equation (\ref{Hz})) the result  
\be
\sqrt{8\pi G}\phi(z)=\frac{2}{3}\sqrt{\frac{3\alpha(1-\Omega_{mo})}{\alpha+3\Omega_{m0}}}ln\Bigg[2(\alpha+3\Omega_{m0})(1+z)^{\frac{3}{2}}+2\sqrt{(\alpha+3\Omega_{m0})^2(1+z)^3+3(1-\Omega_{m0})(\alpha+3\Omega_{m0})}\Bigg].
\label{phiz}
\ee
An addition of the field equations (\ref{friedmann1}) and (\ref{friedmann2}) will now yield 
\be
8\pi G V(z)=\frac{3\alpha H_0^2(1-\Omega_{m0})}{2(\alpha+3)}(1+z)^3+\frac{9H_0^2(1-\Omega_{m0})}{(\alpha+3)}.
\label{Vz}
\ee 
In this expression $z$ can be replaced by $\phi$  using equation (\ref{phiz}) to obtain the potential as a function of $\phi$ as,
\be
8\pi G V(\phi)=\frac{3H_0^2(1-\Omega_{m0})exp(\Phi)}{128(\alpha+3)(\alpha+3\Omega_{m0})^2}+\frac{27H_0^2(1-\Omega_{m0})^3exp(-\Phi)}{2(\alpha+3)}+\frac{9H_0^2(1-\Omega_{m0})(3\alpha+\alpha\Omega_{m0}+12\Omega_{m0})}{4\alpha(\alpha+3)(\alpha+3\Omega_{m0})},
\label{Vphi}
\ee 

where $\Phi=3\sqrt{8\pi G}\sqrt{\frac{\alpha+3\Omega_{m0}}{3\alpha(1-\Omega_{m0})}}\phi$. 
\vskip 1.0cm

%%%%%%%%%%%%%%%%%%%%%%%%%%%
%%%%%%%%%%%%%%%%%%%%%
\begin{figure}[tb]
\begin{center}
\includegraphics[angle=0, width=0.3\textwidth]{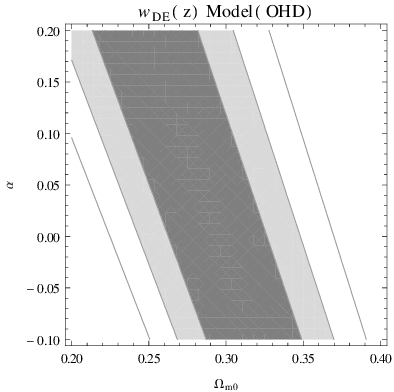}
\includegraphics[angle=0, width=0.3\textwidth]{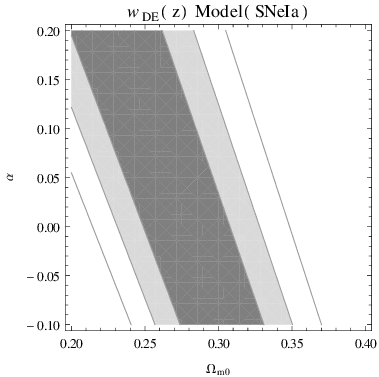}\\
\includegraphics[angle=0, width=0.3\textwidth]{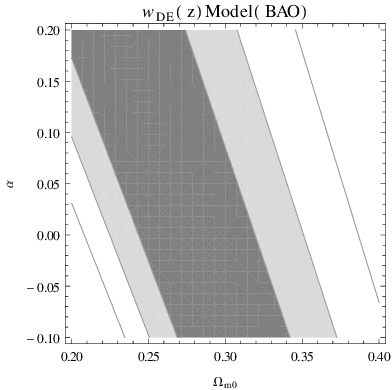}
\includegraphics[angle=0, width=0.3\textwidth]{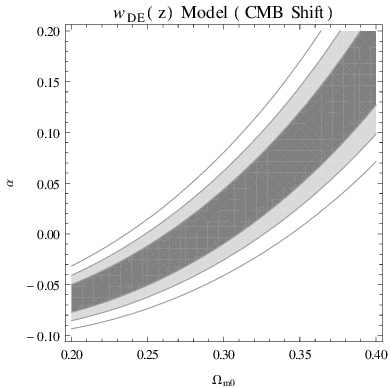}
\end{center}
\caption{{\small Confidence regions on the 2D parameter space obtained for individual data sets. The 1$\sigma$, 2$\sigma$ and 3$\sigma$ confidence regions are presented from inner to outer area. The upper left one is obtained for OHD, upper right is for SNeIa, lower left is for BAO and lower right is for CMBShift data.}}
\label{confidenceregion}
\end{figure}
%%%%%%%%%%%%%%%%%%%%%%

%%%%%%%%%%%%%%%%%%%%%
\begin{figure}[tb]
\begin{center}
\includegraphics[angle=0, width=0.3\textwidth]{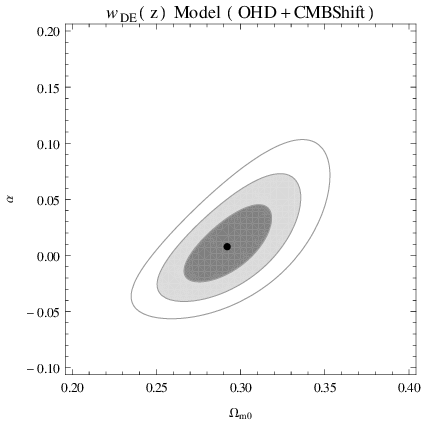}
\includegraphics[angle=0, width=0.3\textwidth]{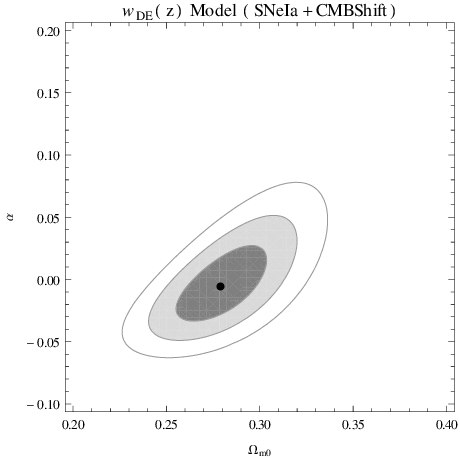}\\
\includegraphics[angle=0, width=0.3\textwidth]{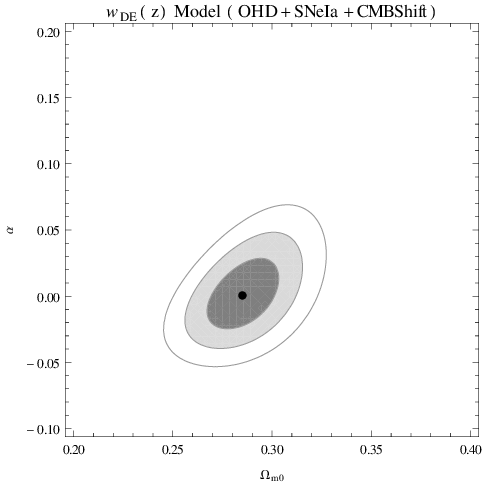}
\includegraphics[angle=0, width=0.3\textwidth]{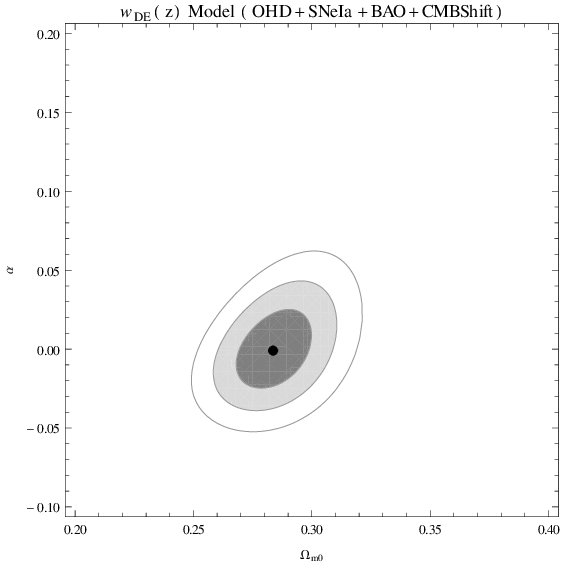}
\end{center}
\caption{{\small Confidence contours on the 2D parameter space obtained for different combinations of the data sets. The 1$\sigma$, 2$\sigma$ and 3$\sigma$ confidence regions are presented from inner to outer area. The central black dots represent the corresponding best fit points. The upper left one is obtained for (OHD+CMBShitf), upper right is for (SNeIa+CMBShft), lower left is for (OHD+SNeIa+CMBShift) and lower right is for (OHD+SNeIa+BAO+CMBShift).}}
\label{contourplots}
\end{figure}
%%%%%%%%%%%%%%%%%%%%%%

\section{Observational constraints  on the parameters}
The essential part of parametric reconstruction is the estimation of the parameter values from the observational data. There are two parameters in the model, the matter density parameter  $\Omega_{m0}$ and the parameter $\alpha$ which is introduced through the expression of $w_{DE}$. Here the observational Hubble data (OHD), type Ia supernova distance modulus data (SNeIa), Baryon Acoustic Oscillation data (BAO) and the CMB shift parameter (CMBShift) data have been used for the statistical analysis. 
\par The observational Hubble data set (OHD) is obtained from the measurement by different groups. Hubble parameter is measured directly from cosmic chromometres and differential age of galaxies in the redshift range $0<z<1.8$ \cite{simonverde,sternjimenz,moresco,zhangc}. Measurement of Hubble parameter at $z=2.3$ \cite{Busca} has also been incorporated in the data set. 
\par The distance modulus ($\mu(z)$) data set from type Ia supernova observations is very widely used one for the analysis of dark energy models. In the present work, the SNeIa data set of Union 2.1 compilation \cite{union21} has been utilized.
\par Baryon Acoustic Oscillation data (BAO) \cite{bao1,bao2,bao3} along with the measurement of {\it comoving sound horizon} at photon decoupling epoch ($z_*=1090.43\pm0.65$) and at photon drag epoch ($z_d=1059.29\pm0.65$) and the estimation of the value of {\it acoustic scale} at decoupling obtained from Planck results \cite{Planck,WangWang} have been incorporated in the statistical analysis.
\par Cosmic Microwave Background (CMB) data, in  the form of a distance prior, namely the CMB shift parameter $R_{\tiny CMBShift}$, estimated from Planck data in \cite{WangWang}, has also been utilized here.  

\par
\begin{figure}[tb]
\begin{center}
\includegraphics[angle=0, width=0.25\textwidth]{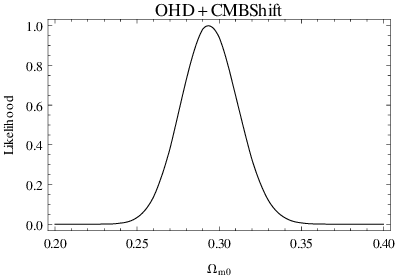}
\includegraphics[angle=0, width=0.25\textwidth]{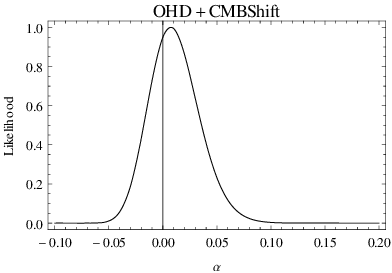}\\
\includegraphics[angle=0, width=0.25\textwidth]{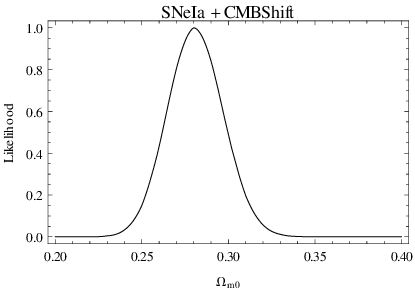}
\includegraphics[angle=0, width=0.25\textwidth]{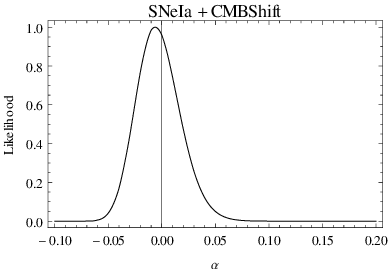}\\
\includegraphics[angle=0, width=0.25\textwidth]{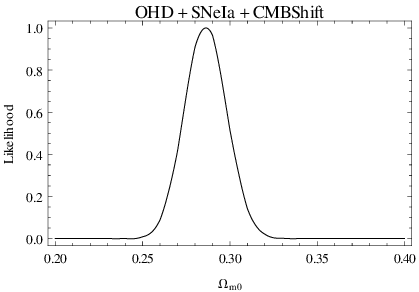}
\includegraphics[angle=0, width=0.25\textwidth]{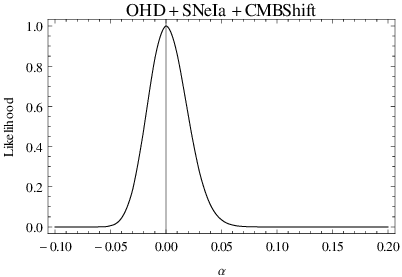}\\
\includegraphics[angle=0, width=0.25\textwidth]{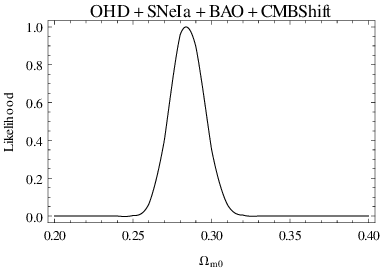}
\includegraphics[angle=0, width=0.25\textwidth]{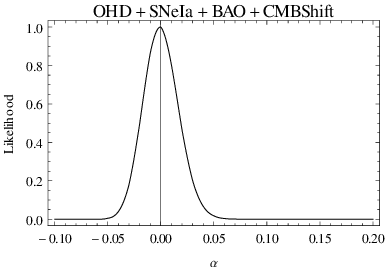}
\end{center}
\caption{{\small Plots of marginalised likelihood functions for different combinations of the data sets. Left panels show the likelihood as a function of $\Omega_{m0}$ and the right panels show the likelihood as function of $\alpha$.}}
\label{likelihood}
\end{figure}

 $\chi^2$-minimization (which is equivalent to the Maximum Likelihood Analysis) technique has been adopted in  the present work for the statistical analysis where $\chi^2$, a function of the set of model parameters $\{\theta\}$, is defined as:
\be
\chi^2(\{\theta\})=\sum_{i=1}^n\frac{(\epsilon_i^{obs}-\epsilon_i^{th}(\{\theta\}))^2}{\sigma_i^2},
\ee
where $\epsilon_i^{th}(\{\theta\})$ is the estimate of the $i^{th}$ data from the model,  $\epsilon_i^{obs}$ stands for the $i^{th}$ observational data and $\sigma_i$ is the error bar associated to the $i^{th}$ data point.  The Likelihood function is defined as:
\be
{\mathcal L}=\exp{\Big(-\frac{\chi^2(\{\theta\})}{2}\Big)}.
\ee
Statistical analysis has been carried out for each data set individually as well as for different combinations of the them. The statistical analysis for combination of data sets has been done by adding up the $\chi^2$ functions of each data sets which are taken into account for that combination, i.e. $\chi^2_{total}=\sum\chi^2$. Figure \ref{confidenceregion} shows the confidence regions on 2D parameter space obtained for each data sets individually. The confidence regions on the parameter space (figure \ref{confidenceregion}) are not closed and thus the parameters $\alpha$ and $\Omega_{m0}$ are not at all well constrained and wide ranges of them are allowed for each data set. Furthermore, the CMB shift parameter data indicates a qualitatively  different confidence region on the parameter space. Figure \ref{contourplots} presents the confidence contours on the parameter space for various combinations of the data sets with CMB shift parameter data being common in all the combinations. All the combinations used put effectively tighter constraints on the parameters making the model more precise.
Table \ref{tableCombinedanalysis} contains the best fit values of the parameters $\alpha$ and $\Omega_{m0}$ along with the allowed variation in the 1$\sigma$ error bar. The best fit values are obtained by the usual $\chi^2$ minimization technique.   Figure \ref{likelihood} presents the marginalised likelihood functions. The likelihood function plots are well fitted to Gaussian distribution for both the parameters as arguments.

%%%%%%%%%%%%%%%%%%%%%%%%%
\begin{table}[h!]
\begin{center}
\resizebox{0.8\textwidth}{!}{  
\begin{tabular}{ |c ||c |c |c |} 
 \hline
  Data  & $\chi^2_{min}/d.o.f.$ & Parameters \\ 
 \hline
 \hline
  {\small OHD+CMBShift} &  13.34/26 & $\Omega_{m0}=0.292\pm0.012$, $\alpha=0.0078\pm0.0160$\\ 
 \hline
  {\small SNeIa+CMBShift} &  562.23/577 & $\Omega_{m0}=0.279\pm0.012$, $\alpha=-0.0056\pm0.0156$\\ 
  \hline
  {\small SNeIa+OHD+CMBShift} & 575.89/603 &  $\Omega_{m0}=0.285\pm0.008$, $\alpha=0.0005\pm0.0124$\\ 
  \hline
  {\small SNeIa+OHD+BAO+CMBShift} &  578.04/606 & $\Omega_{m0}=0.284\pm0.007$, $\alpha=-0.0009\pm0.0117$\\ 
  \hline
\end{tabular}
}
\end{center}
\caption{{\small Results of the statistical analysis. The reduced $\chi^2$ i.e. $\chi^2_{min}/d.o.f.$ where $d.o.f.$ is the number of degrees of freedom of that $\chi^2$ distribution, the best fit values of the parameters along with the 1$\sigma$ error bar obtained for different combinations of the data sets are presented.}}
\label{tableCombinedanalysis}
\end{table}
%%%%%%%%%%%%%%%%%%%%%%%%%%%%%%%%%%%

%%%%%%%%%%%%%%%%%%%%%
%%%%%%%%%%%%%%%%%%%%%%

%%%%%%%%%%%%%%%%%%%%%%%%%%%%%%%%%%%%%%%
\begin{figure}[htb]
\begin{center}
\includegraphics[angle=0, width=0.3\textwidth]{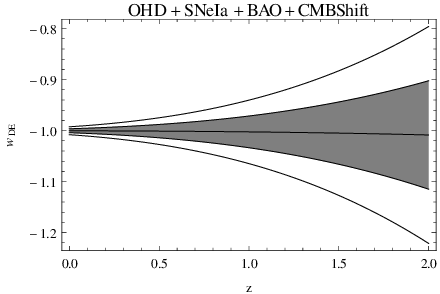}
\includegraphics[angle=0, width=0.3\textwidth]{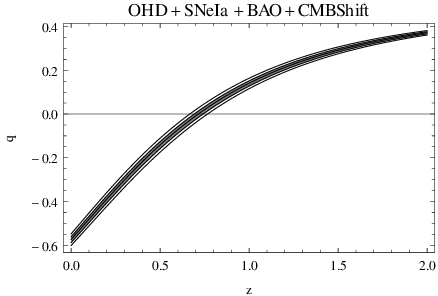}\\
\includegraphics[angle=0, width=0.3\textwidth]{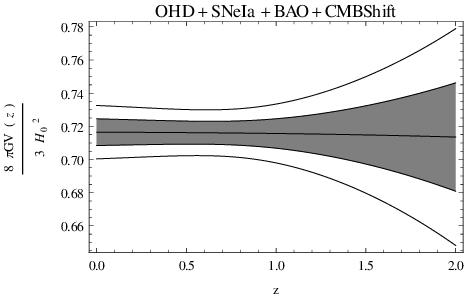}
\includegraphics[angle=0, width=0.32\textwidth]{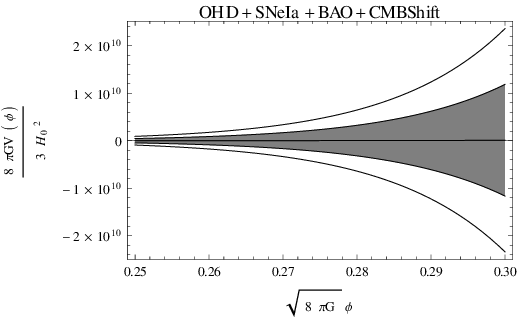}
\end{center}
\caption{{\small The upper panels show the behaviour of dark energy equation of state parameter $w_{DE}(z)$ (left) and the deceleration parameter $q(z)$ (right) as a function of redshift $z$ within 1$\sigma$ and 2$\sigma$ confidence levels with the central black line representing the best fit curve. The lower panels show the quintessence potential as a function of redshift $z$ (left) and also as a function of the quintessence scalar field   (right) for 1$\sigma$ and 2$\sigma$ confidence level with the best fit curve being given by the central dark line.}}
\label{wzqzVzVphi}
\end{figure}
%%%%%%%%%%%%%%%%%%%%%%

\vskip 2.8cm

\section{Discussion}
The present work presents a quintessence model where a dark energy equation of state parameter $w_{DE}$, which is chosen as a one parameter function of $z$, is reconstructed from the observational data.  

\par From figure \ref{confidenceregion}, it is clear that the model is not well constrained when the statistical analysis is carried out using the data sets individually. But much tighter constraints on the parameters can be obtained using proper combination of the data sets (figure \ref{contourplots}). The last row in table \ref{tableCombinedanalysis} shows the best fit values of the model parameter $\alpha$ (which describes the equation of state given by equation (\ref{wz})) and the matter density parameter $\Omega_{m0}$,  obtained for the combination of SNeIa, OHD, BAO and CMB shift parameter data. The values are given as $\alpha=-0.0009\pm0.0117$ and  $\Omega_{m0}=0.284\pm0.007$ in 1$\sigma$ confidence region. The best fit value of $\alpha$ is negative, but very close to zero. That means the proposed model is very  close to $\Lambda$CDM with a tendency towards favouring the phantom nature of dark energy. The present value of dark energy equation of state parameter ($w_0$) is constrained to be $ w_0=-1.000\pm0.004$ at 1$\sigma$ confidence level by the present reconstruction.
\par The upper right panel of figure 4 shows that the deceleration parameter $q$ starts positive for a higher $z$, and attains negative value near $z=0$ with a signature flip between $z=0.6$ and $0.8$. This is consistent with the present observation \cite{ratra}. 
\par The reconstructed quintessence potential is shown in the lower panel figure \ref{wzqzVzVphi} corresponding to the combinations of the data sets used. The left panel, which depicts $V=V(z)$, clearly indicates  that $V(z)$ remains almost flat. So one can say that the potential is a ``freezing" potential as opposed to a thawing one (see for example, the work of Caldwell and Linder\cite{cald} and that of Scherrer and Sen \cite{scherrer}). 
\par The potential is shown in the lower right panel of figure \ref{wzqzVzVphi} and the analytic form is given in equation (\ref{Vphi}).  A similar potential had already been discussed by Sen and Sethi \cite{ss}. The potential obtained in the present work in the form $V(\phi)=V_1e^{\lambda\phi}+V_2e^{-\lambda\phi}+V_0$ where $V_1$, $V_2$,$V_0$ and $\lambda$ are constants, is a generalization of the potential given by Sen and Sethi, where $V_1=V_2$.  The requirement of $V_{1}= V_{2}$ would yield, from equation (\ref{Vphi}), the condition
\be
{\Omega}_{m0} = \frac{-(\alpha - 3) + \sqrt{(\alpha - 3)^{2} + 12(\alpha \mp \frac{1}{24})}}{6}.
\label{v1v2}
\ee
If we take realistic values of ${\Omega}_{m0}$ between $\frac{1}{4}$ and $\frac{1}{3}$, the value of $\alpha$ will lie between $-\frac{25}{36}$ and $-\frac{17}{16}$ leading to the values of $w_{DE}$ at $z=0$ between -1.301 and -1.548 respectively, well into the phantom regime of $w_{DE} < -1$. But this is out of 2$\sigma$ error bar of $w_{DE}$ at $z=0$ of the model presented in this work. Thus although the present model allows for a phantom regime, the Sen and Sethi model is not favoured.
\par A recent analysis \cite{XiaJQ} using CMB temperature anisotropy and polarization data, along with other non-CMB data, estimates the values of the parameters as $\Omega_{m0}=0.293\pm0.013$ at 1$\sigma$ confidence level for the $\Lambda$CDM model where $w_0=-1$ (not a parameter) and $\Omega_{m0}=0.270\pm0.014, w_0=-1.167\pm0.061$ at 1$\sigma$ confidence level for $w$CDM model. The present model is inclined towards the $\Lambda$CDM model, and the value of $\Omega_{m0}$ remains in between the values obtained for $\Lambda$CDM and $w$CDM (within 1$\sigma$ confidence level of both the models). 

\par A study of different parameterizations of dark energy equation of state by Hazra {\it et al} \cite{hazra} has  obtained the parameter values in various cases. For example, Chevallier-Polarski-Linder (CPL) parametrization \cite{cpl1,cpl2} yields $\Omega_{m0}=0.307_{-0.046}^{+0.041}, w_0=-1.005_{-0.17}^{+0.15}$, Scherrer and Sen (SS) parameterization \cite{scherrer} yields $\Omega_{m0}=0.283_{-0.030}^{+0.028}, w_0=-1.14_{-0.09}^{+0.08}$ and generalized Chaplygin gas (GCG) parameterization \cite{gcg} shows  $\Omega_{m0}=0.32_{-0.012}^{+0.013}, w_0=-0.957_{\tiny non-phantom}^{+0.007}$. All the estimates are at a  1$\sigma$ confidence level. Hence the present reconstruction is consistent with the CPL parameterization at 1$\sigma$ confidence level. The SS parameterization requires a slightly  lower value of $w_0$ (out of 1$\sigma$ error bar) but the value of $\Omega_{m0}$ is highly consistent. The non phantom prior assumption of GCG parameterization is not in agreement with the present model, but the lower bound of 1$\sigma$ error bar for GCG parameterization is within the 1$\sigma$ confidence region of the present model. 
\par A reconstruction of quintessence potential described by a polynominal serise constrains the present value of dark energy equation of state $w_0=-0.978_{-0.031}^{+0.032}$ \cite{hutererpeiris}. This is within the 1$\sigma$ error bar of the present model. 
\par It deserves mention that systematic uncertainties of observations might have its imprints on the results of these analyses. For instance, the colour-luminosity parameter might depend on the redshift, and hence affect the magnitude in the analysis of Supernova data \cite{wang}. We also refer to the analyses of Rubin {\it et al} \cite{rubin} and Shafer and Huterer \cite{shaferhuterer} for some very recent development in connection with the systematics.

\vskip 1.50cm

{\bf Acknowledgment:} The authors would like to thank Anjan Ananda Sen, Sumit Kumar and Md. Wali Hossain for their valuable help regarding the statistical analysis. They would also like to thank the anonymous referee, whose suggestions led to a substantial improvement of the paper.

\end{document}